\documentstyle[psfig]{cupconf}

\ifoldfss
\else
  \ifnfssone
    \newmathalphabet{\mathit}
      \addtoversion{normal}{\mathit}{cmr}{m}{it}
      \addtoversion{bold}{\mathit}{cmr}{bx}{it}
    \newmathalphabet{\mathcal}
      \addtoversion{normal}{\mathcal}{cmsy}{m}{n}
    \else
    \ifnfsstwo
    \fi
  \fi
\fi

\def\ri{r_{I}}
\def\ro{r_{O}}
\def\aproxgt{\mathrel{%
        \rlap{\raise 0.511ex \hbox{$>$}}{\lower 0.511ex \hbox{$\sim$}}}}
\def\aproxlt{\mathrel{%
        \rlap{\raise 0.511ex \hbox{$<$}}{\lower 0.511ex \hbox{$\sim$}}}}

\def\upi{\pi} % to be replaced with upright Greek character
\def\hexnumber#1{\ifcase#1 0\or1\or2\or3\or4\or5\or6\or7\or8\or9\or
 A\or B\or C\or D\or E\or F\fi }

\makeatletter
\ifx\CUP@mtlplain@loaded\undefined
\else
\fi
\makeatother
 \makeatletter
 \ifx\CUP@mtlplain@loaded\undefined
   \font\tenbmi=cmmib10 at 10pt
   \font\sevenbmi=cmmib10 at 7pt
   \font\fivebmi=cmmib10 at 5pt

   \newfam\bmifam
   \textfont\bmifam=\tenbmi
   \scriptfont\bmifam=\sevenbmi
   \scriptscriptfont\bmifam=\fivebmi
   
 \fi
 \makeatother
\ifnfsstwo

\fi
\ifnfssone

\fi
\ifoldfss

\fi

\mathchardef\varLambda="0103

\makeatletter
\ifx\CUP@mtlplain@loaded\undefined
\else
\fi
\makeatother
\makeatletter
\ifx\CUP@mtlplain@loaded\undefined
  \font\tenbms=cmbsy10
  \font\sevenbms=cmbsy10 at 7pt
  \font\fivebms=cmbsy10 at 5pt
  \newfam\bmsfam
  \textfont\bmsfam=\tenbms
  \scriptfont\bmsfam=\sevenbms
  \scriptscriptfont\bmsfam=\fivebms

  \edef\bsy@{\hexnumber\bmsfam}
  \mathchardef\bnabla="0\bsy@72
\fi
\makeatother
\title[Theory of Black Hole Accretion Discs]{Stable
Oscillations of Black Hole \\ Accretion Discs}

\author[M.A. Nowak \& D.E. Lehr]%
{M\ls I\ls C\ls H\ls A\ls E\ls L\ns A.\ns N\ls O\ls W\ls A\ls K$^1$ 
\and \ns D.\ns E.\ns L\ls E\ls H\ls R$^2$}

\affiliation{$^1$JILA, Campus Box 440, Boulder, CO 80309-0440, USA\\[\affilskip]
$^2$Department of Physics, Stanford University, Stanford, CA 94305-0460, USA}

\begin{document}
\ifnfssone
\else
  \ifnfsstwo
  \else
    \ifoldfss
      \let\mathcal\cal
      \let\mathrm\rm
      \let\mathsf\sf
    \fi
  \fi
\fi

\maketitle

\begin{abstract}
The study of stable accretion disc oscillations relevant to black hole
candidate (BHC) systems dates back over twenty years.  Prior work has
focused on both unstable and (potentially) stable disc oscillations.  The
former has often been suspected of being the underlying cause for the
observed broad-band variability in BHC, whereas the latter has had little
observational motivation until quite recently.  In this article, we review
both the observations and theory of (predominantly) stable oscillations in
BHC systems.  We discuss how variability, both broad-band and
quasi-periodic, is characterized in BHC. We review previous claims of low
frequency features in BHC, and we discuss the recent observational evidence
for stable, high frequency oscillations in so-called `galactic
microquasars'.  As a potential explanation for the latter observations, we
concentrate on a class of theories-- with a rich history of study-- that we
call `diskoseismology'.  We also discuss other recent alternative theories,
namely Lense-Thirring precession of tilted rings near the disc inner edge.
We discuss the advantages and disadvantages of each of these theories, and
discuss possible future directions for study.
\end{abstract}

\firstsection % if your document starts with a section,
\section{Overview of Black Hole States}

Galactic X-ray sources are typically identified as black hole
candidates (BHC) if they have measured mass functions indicating a
compact object with $M \aproxgt 3~ M_\odot$, or if their high energy
spectra ($\sim 1$ keV $-$ $10$ MeV) and temporal variability ($\sim
10^{-3} - 10^2$ Hz) are similar to other BHC.  A review of the general
observations, a number of theoretical models, plus individual
descriptions of  approximately twenty galactic BHC can be
found in \cite{tanaka95}.  A review of the timing analyses can
be found in \cite{vanderklis94} and \cite{vanderklis95}.  (These
latter reviews attempt to link the energy spectra and timing
observations, and they draw analogies to similar observations of
neutron stars in low mass X-ray binaries.)  A review of a subset of BHC
with reasonably well-determined mass functions can be found in \cite{nowak95}.

The energy spectra of BHC have been historically labelled based upon
observations of the soft X-ray band ($\sim 2 - 10$ keV).  Intense,
quasi-thermal flux is referred to as the ``high'' state.  Non-thermal flux
in this band, typically a power law with a photon index\footnote{Photon
index here and throughout shall refer to the photon count rate, such that
photon index $\Gamma$ implies \# photons/keV/s/cm$^2$ $\propto ~
E^{-\Gamma}$, where $E$ is the photon energy.} of $\sim 1.7$, indicates
that the BHC is in the ``off'' state (for extremely low intensity flux) or
``low'' state (for moderate intensity flux).  The high state tends to have
little variability, with a root mean square (rms) variability
(cf. \S\ref{sec:var}) of a few percent, whereas the low state tends to have
an rms variability of several tens of percent.  If a black hole candidate
has a quasi-thermal soft X-ray component and significant high energy
emission--- occasionally modeled as a power law with a photon index $\sim
2.5$--- it is said to be in the ``very high'' state
[cf. \cite{miyamoto91}].  The rms variability of the very high state tends
to be greater than that of the high state, but less than that of the low
state [cf. \cite{canonicalI}, \cite{canonicalII}].  Note that the labels
``off'', ``low'', ``high'', and ``very high'' are purely qualitative in
nature, but historically have been in popular use.  No universally agreed
upon quantitative definition for these states exist.  More recently, the
terms ``hard state'' and ``soft state'' have begun to be used in place of
``low state'' and ``high/very high state'', respectively.

\begin{figure} 
\vspace{0.1 true in}
\centerline{\psfig{figure=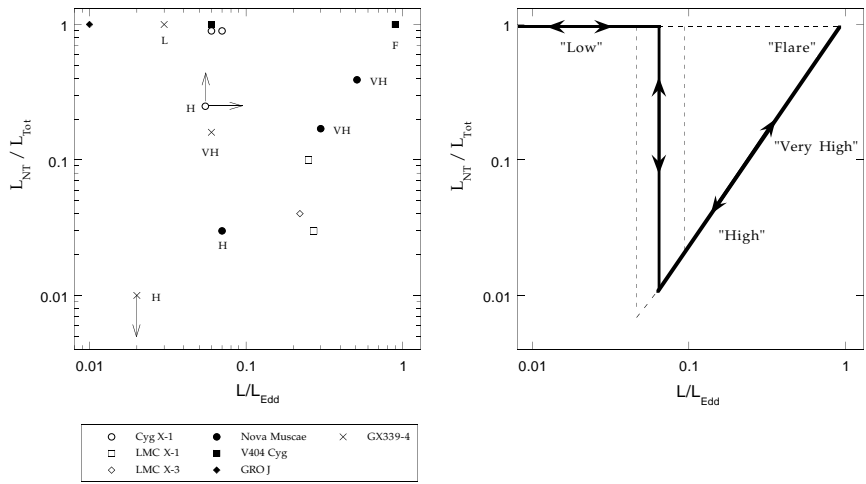,width=0.85\textwidth}}
  \caption{{\it Left:} The ratio of luminosity in a power
  law tail to total luminosity ($L_{\rm NT}/L_{\rm Tot}$) versus the ratio of
  total luminosity to Eddington luminosity ($L/L_{\rm Edd}$) for black
  holes with well-determined mass functions.  (Note:  error bars are {\it
  not} plotted.) {\it Right:} Hypothetical evolutionary tracks for changes
  in BHC luminosity.  [Figure adapted from \cite{nowak95}.]}
  \label{fig:lum}
\end{figure} 

Figure \ref{fig:lum} depicts this state behavior for several BHC with
reasonably well-determined mass functions [cf. \cite{nowak95}].  This
figure shows the fraction of total luminosity in a power law-like tail as a
function of the BHC's fractional Eddington luminosity.  BHC tend to be
dominated by a power law tail for luminosities below $\sim 10\% ~L_{\rm
Edd}$, be completely dominated by a quasi-thermal, disc-like component near
$\sim 10\% ~L_{\rm Edd}$, and then begin to show a power law-like tail for
$L \aproxgt 10\%~L_{\rm Edd}$.  Above $10\%~L_{\rm Edd}$, the fraction of
total luminosity in this power law tail (which typically has photon index
$\Gamma \sim 2.5$) tends to increase with fractional Eddington luminosity.
There are, of course, exceptions to the trends shown in 
Fig. \ref{fig:lum}; however, some transient BHC, such as Nova Muscae, have
followed this trend very closely [cf. \cite{miyamoto94}].  Historically,
the soft flux component has been associated with a ``classical'' accretion
disc, as discussed by \cite{shakura73}.

\section{Measuring Variability}\label{sec:var}

\begin{figure} 
\vspace{0.1 true in}
\centerline{\psfig{figure=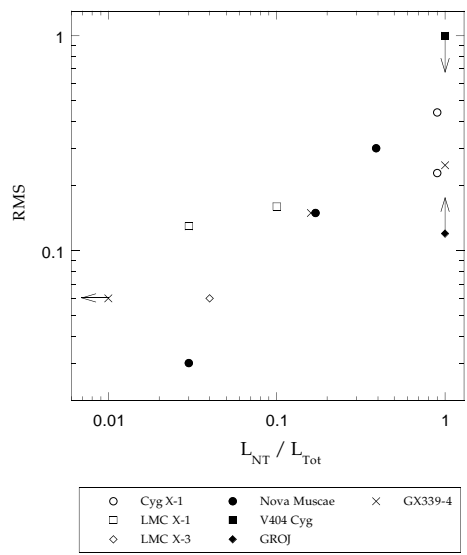,width=0.45\textwidth}}
  \caption{RMS variability versus ratio of luminosity in a power law tail
  to total luminosity (cf. Fig. \ref{fig:lum}).  (Consistent energy bandpasses
  were not used for all depicted data points, although they predominantly
  represent a $\sim 2-35$ keV bandpass.  Again, error bars are {\it not}
  plotted.) [Figure adapted from \cite{nowak95}.]}	
  \label{fig:var}
\end{figure}

The states of BHC have been further distinguished by their variability
behavior, as discussed by \cite{vanderklis94}, \cite{vanderklis95},
\cite{canonicalI}, and Miyamoto, et al. (1993).  As shown in Figure
\ref{fig:var}, the greater the fraction of luminosity in a hard tail, the
greater the amplitude of the observed variability.  Miyamoto, et al. (1992)
and Miyamoto, et al. (1993) further point out that states appear to have
``canonical'' variability behaviour not only in terms of amplitude, but
also in terms of frequency distribution.  This frequency distribution is
typically described via Fourier transfrom techniques
[cf. \cite{vanderklis94}], which we briefly describe below.

The key assumption in applying Fourier transform techniques is the
assumption that the lightcurve represents a statistically stationary process
[cf. \cite{davenport}].  Effectively, this means that one assumes that
individual data segements from the lightcurve are a good representation of
an ensemble of {\it statistically} identical lightcurves.  It is from these
individual lightcurve segments that one forms the power spectral density
(PSD) [cf. \cite{vanderklis94}].

The PSD is calculated by dividing the lightcurve into segments of equal
length and then taking a Fast Fourier Transform (FFT, i.e. discrete
transform, cf. Press, et al. 1992) of each data segment.  The squared
amplitude of each individual FFT is then averaged together.  Often one also
averages over Fourier frequency bins.  (Typically one chooses a scheme such
that the number of frequency bins averaged over is $\propto f$, where f is
the Fourier frequency.) This yields the resulting PSD for each lightcurve.

There are a number of different possible normalizations for the PSD;
however, in this work we shall only refer to the normalization of
\cite{canonicalI}.  For this normalization, the integral of the PSD over
positive Fourier frequency yields the {\it square} of the total {\it root
mean square (rms) amplitude} of the variability [\cite{canonicalI}]. That is,
the PSD integrated over positive frequency yields $(\langle x^2\rangle -
\langle x\rangle^2 )/\langle x \rangle^2$, where $x$ is the photon count
rate [cf. \cite{canonicalI}, \cite{canonicalII}].  The rms can be
calculated over specific Fourier frequency and energy spectral bandpasses.
For a given narrow Fourier frequency interval, the rms is (to within a
factor of $\sqrt{2}$) the fractional amount by which the lightcurve is
sinusoidally modulated in that given Fourier frequency interval.

It is this measure of rms variability, typically calculated over frequency
intervals of $\sim 10^{-3}-10^{2}$ Hz in BHC, that leads to the
characterization of power law tail-dominated sources (i.e. low and very
high state sources) as being the most variable BHC.  As a function of
Fourier frequency, the PSD of all states can usually be described by
(possibly multiple) broken power laws.  However, occasionally superimposed
on top of this broad band variability are narrow--- compared to the the
broad band PSD--- features which are referred to as quasi-periodic
oscillations (QPO).  The properties of these features are
the focus of the next section.

\section{Quasi-periodic Variability observed in BHC}

A quasi-periodic oscillation is usually taken to be {\it any} feature in
the PSD that is well-fit by a Lorentzian of the form:
\begin{equation}
PSD(f) ~=~ {{R^2 ~Q ~f_{QPO}}\over{{\upi \left [~ Q^2 (f_{QPO} - f )^2 ~+~
f_{QPO}^2 ~\right ]}}} ~~.
\label{eq:lorentz}
\end{equation}
As written in eq. \ref{eq:lorentz}, this functional fit is relevant to the
normalization of Miyamoto, et al. (1992).  In the above, $R$ is the
fractional rms, $f_{QPO}$ is the QPO frequency, and $Q$ is the mode
`quality factor'.  If $\Delta f_{QPO}$ is the full-width half maximum of
the QPO feature, then $Q \approx f_{QPO}/\Delta f_{QPO}$.

For $Q \aproxlt 3$, it is difficult to attribute the QPO to a discrete
feature, as opposed to being merely an extension of the broad-band
variability.  For $Q \aproxgt 10$, the QPO is usually quite apparent in the
PSD, and may be attributable to a discrete process in the BHC system.  The
question then arises of whether or not such a QPO feature represents a
quasi-stable mode in an accretion disc.  If the QPO feature is attributable
to a mode or set of modes, then there are many possible mechanisms for
generating a finite feature width in the PSD [cf. \cite{vanderklis94}].
Among these possibilities are: multiple modes distributed over frequencies
$f_{QPO} \pm \Delta f_{QPO}$, a driven-damped mode with damping time scale
$\tau_D \approx Q f_{QPO}^{-1}$, a mode that executes a random walk in
phase of $\sim 2 \upi$ on time scales of order $\tau_D$, etc.  Rarely does
one have the statistics to determine which of these processes is
responsible for the $Q$ of the $QPO$.

Prior to the launch of the {\it Rossi X-ray Timing Explorer (RXTE)},
observations of QPO in BHC were made with {\it Ginga}, {\it BATSE}, and
{\it Granat/SIGMA}.  All of these observed QPO were at frequencies
$\aproxlt 10$ Hz, although the above instruments were for the most part
limited to $\aproxlt 60$ Hz.  In addition, the rms variability of these
features were all $\aproxlt 10\%$, with $Q \aproxlt {\cal
O}(10)$.  Below, we review some of the historical observations of black
hole QPO. 

\subsection{Historical Observations}\label{sec:obshist}

Reviews of some of the observed QPO features in BHC can be found in
\cite{nowak95} and \cite{tanaka95}.  Here we consider separately
low state and high/very high state observations of BHC .

In the low state, Cygnus X--1 has shown QPO with rms as high as $15\%$
[cf. Kouveliotou, et al. (1992a), Ubertini, et al. (1994), Vikhlinin, et
al. (1994)]. The features were seen to have frequencies $\sim 0.04$ and
$0.07$ Hz.  However, they appeared more as broad peaks in the otherwise
broken power law PSD, and did not appear as discrete, narrow features.
Similar features, with frequencies $0.04$ and $0.2$ Hz were seen in the
hard state of the X-ray transient GRO~J0422+32 (a.k.a. Nova Persei)
[Kouveliotou, et al. (1992b), Sunyaev, et al. (1992),
Sunyaev, et al. (1993)].

The source GX~339--4, which has been observed to transit through soft and
hard X-ray states, has shown a relatively narrow [$Q \sim {\cal O}(10)$]
QPO at $0.8$ Hz [\cite{grebenev91}] during its hard state.  On the other
hand, in its very high state GX~339--4 has also shown a narrow QPO at $6$
Hz [\cite{miyamoto91}].  Similar to this feature, the X-ray transient Nova
Muscae has shown QPO in the range $3-8$ Hz during its very high state
[\cite{kitamoto92}]. 

In the soft state of LMC X-1 (which is the only state that has been
observed in this source), there has been a claim of a $0.08$ Hz feature
with rms $\sim 4\%$ [Ebisawa, Mitsuda, \& Inoue (1989)].  However, the
count rates for these observations were extremely low, and it is unclear
whether or not the putative variability detections described by Ebisawa,
Mitsuda, \& Inoue (1989) were below the effective Poisson noise limit.
More recently, \cite{cui97} has found evidence (based upon {\it RXTE}
observations) of a $3-12$ Hz feature in the soft state of Cygnus X--1.  We
note, however, that this feature is very broad and is difficult to
associate with a discrete feature distinct from the observed broad-band
variability.

Prior to the recent {\it RXTE} observations, a fairly clean division of QPO
frequencies occurs if one ignores the $0.08$ Hz QPO attributed to LMC
X--1.  Low frequency QPO ($\aproxlt 1$ Hz) are seen in low/hard states, while
high frequency QPO ($\sim 6$ Hz) are seen in very high/soft states.  Again,
two questions arise.  First, are these QPO associated with modes distinct
from the broad-band variability?  Second, if the QPO are distinct modes,
are they associated with an accretion disc or with another component (such as a
corona) in the BHC system?  In the next subsection, we discuss a set of 
observed features in two BHC that may indeed be attributable to an
accretion disc.
 
\subsection{QPO in `Microquasars'}\label{sec:micro}

Recently, two rather unusual and dramatic X-ray transient galactic BHC have
been observed: GRS~1915+105 [Mirabel \& Rodriguez (1994), Morgan,
Remillard, \& Greiner (1997), Chen, Swank, \& Taam (1997), Taam, Chen, \&
Swank (1997)] and GRO~J1655-40 [Hjellming \& Rupen (1995), Remillard et
al. (1997)].  These sources were unusual in that they showed powerful,
highly relativistic radio jets [Mirabel \& Rodriguez (1994), Hjellming \&
Rupen (1995)]; hence, they have been dubbed `microquasars'.  Furthermore,
both have shown dramatic X-ray variability.

As observed by {\it RXTE}, GRS~1915+105 has shown X-ray count rates up to
$10^5$ cts/sec, which, given a distance of $\aproxgt 12$ kpc [Mirabel \&
Rodriguez (1994)], indicates a peak luminosity of over $10^{40}$ erg/s
[Morgan, Remillard, \& Greiner (1997), Chen, Taam, \& Swank (1997)].  As
discussed in Morgan, Remillard, \& Greiner (1997), GRS~1915+105 has also
shown intense variability patterns, with the luminosity changing by factors
of several in only a few seconds.  Amidst this spectral behavior, a host of
different QPO features have been observed, ranging in frequencies from
$\sim 0.1-10$ Hz [Morgan, Remillard, \& Greiner (1987), Chen, Swank, \&
Taam (1997), Taam, Chen, \& Swank (1997)].  Many of these QPOs have
multiple harmonics, and both correlations and anti-correlations with source
luminosity have been observed.

However, among the various QPO features seen, a high frequency QPO at 67
Hz has stood out because it apparently does not appreciably vary in
frequency [Morgan, Remillard, \& Greiner (1997)].  During the first epoch
that this QPO feature was observed, it was relatively narrow ($Q\sim20$),
weak (rms variability $\sim 0.3-1.6\%$), and varied in frequency by  $<3\%$,
despite factors of $\sim 2$ variations in source luminosity.  When viewed
in restricted bandpasses the rms variability was seen to increase with
energy, with a maximum rms variability $\sim 6\%$ in the highest energy
bandpass ($\sim 10-20$ keV).  During subsequent observations similar
features were observed, always at $67\pm2$ Hz [even on occasions when no
soft, excess flux above a power law tail was observed in the spectrum;
Morgan 1997, Private Communication].

A high frequency QPO feature has also been observed in GRO~J1655-40, with
frequency 300 Hz [Remillard, et al. (1997)].  However, this QPO was only
detected once, and then only by summing several, individual observations.
It was therefore difficult to place limits on the stability of this
feature's frequency.  Like the 67 Hz feature in GRS~1915+105, this feature
was weak (rms variability $\sim 0.8\%$).  Furthermore, it was detectable
only in the ``hardest'' spectra ($\sim 10-20$ keV).  The 300 Hz feature in
GRO~J1655-40 was somewhat broader than the 67 Hz feature in GRS~1915+105.
It is also notable in that a very accurate mass determination,
$7\pm0.2~M_\odot$, has been made for the compact object in GRO~J1655-40
[Orosz \& Bailyn (1997)].

As we will discuss in \S\ref{sec:theory}, a number of
properties displayed by these features, especially the 67 Hz feature in
GRS~1915+105, are what one expects for stable, global oscillations of
accretion discs.  We will concentrate on theories of the oscillation of the
innermost disc regions as potential explanations of these observations.

\subsection{QPO in Active Galactic Nuclei}

As discussed in \S\ref{sec:char}, the characteristic time scales are one
hundred thousand to one hundred million times longer in Active Galactic
Nuclei (AGN) as compared to galactic BHC.  Thus, if a $0.1$~s QPO period is
typical of a galactic black hole, then we might expect a 3 hour to
year-long period to be typical for AGN.  Such long time scales are very
difficult for current missions to detect, as the shorter period is on the
order of the {\it total} duration of a typical observation, and the longer
period requires many dedicated observations.  However, one AGN,
IRAS18325--5926, has shown strong evidence for a $5.5 \times 10^4$ s
periodicity that was observed for 9 cycles [\cite{iwasawa98}].  Over these
few cycles, there was no evidence for any strong change in phase, and the
amplitude of the modulations was on the order of $10\%$.  This is a
somewhat stronger amplitude than observed for the $67$ Hz and and $300$ Hz
features described above, and therefore are difficult to explain with the
models described in \S\ref{sec:theory}.  Further (longer) observations are
required to determine to what extent this AGN QPO is or is not similar to
the QPO seen in the microquasars.

\section{Characteristic Accretion Disc Frequencies and
Time Scales}\label{sec:char} 

The fact that the 67 Hz QPO feature seen in GRS~1915+105 does not appear to
vary in frequency as a function of luminosity suggests that this feature is
tied to a fundamental frequency or time scale in the BHC system.  As this
frequency is apparently independent of accretion rate (i.e. luminosity),
the relevant time scales are likely to be gravitational ones.  The two most
important parameters for the gravitational time scales of BHC are the black
hole mass, $M$, and the angular momentum, $J$.  We shall usually normalize
the black hole mass to a solar mass. We will characterize the black hole
angular momentum by the dimensionless parameter $a \equiv cJ/GM^2$, where
$c$ is the speed of light and $G$ is the gravitational constant.  For a
Schwarzchild black hole, $a = 0$, while for a Kerr hole $a < 1$.  Below we
discuss the frequencies most relevant to orbits near the equatorial plane
of a (possibly spinning) black hole.  These frequencies are those most
relevant to a thin accretion disc\footnote{We define the equatorial plane
to be the plane perpendicular to the black hole angular momentum axis that
passes through the center of the black hole.  We shall consider a
cylindrical coordinate system ($r,\phi,z$) set up on this plane, with $z=0$
being in the plane.}.

In what follows, we shall set $c=G=1$. Furthermore, physical length scales
will be normalized to $GM/c^2$ ($\approx15$ km for $M=10~M_\odot$), and
frequencies will be normalized to $c^3/GM$ ($\approx3.2 \times 10^3$ Hz for $M =
10~M_\odot$).  In general, characteristic gravitational length scales
increase linearly with $M$, whereas characteristic time scales decrease
linearly with $M$.  Therefore, whereas 1 ms may be a relevant time scale for
a galactic black hole, days or months might be the relevant time scale for a
massive black hole at the center of an AGN.  There will be four main
frequencies of concern to us, each a function of radius $r$: the Keplerian
(i.e. orbital) frequency, the radial epicyclic frequency, the vertical
epicyclic frequency, and the Lense-Thirring precession frequency.  We
describe each of these in turn below.

The Keplerian frequency is the frequency with which a free particle
azimuthally orbits the black hole.  This frequency, to a viewer observing
at infinity, is given by:
\begin{equation}
\Omega ~=~ \left ( ~ r^{3/2} ~+~ a ~ \right )^{-1}
\label{eq:omega}
\end{equation}
For a $10~M_\odot$ Schwarzchild black hole, this is approximately 220 Hz at
$r=6$.

The radial epicyclic frequency is the frequency at which a free particle
oscillates about its original circular orbit if it is given a radial
perturbation.  In classical mechanics, the square of the radial epicyclic
frequency is given by $\kappa^2 = 4 \Omega^2 + r \partial \Omega^2/\partial
r$, and is exactly equal to $\Omega^2$ for Keplerian orbits in an $r^{-1}$
potential\footnote{This is just the statement that a radial perturbation
sends a circular orbit into an elliptical one, with an identical orbital
period, for Newtonian free-particle orbits in an $r^{-1}$ potential.}
[cf. \cite{binney}].  In General Relativity, there is a minimum radius for
which an orbit is stable to radial perturbations.  The innermost marginally
stable orbit is at $r=6$ for $a=0$, and moves inward as $a$ increases.  The
epicyclic frequency is given by
\begin{equation}
\kappa^2 ~=~ \Omega^2 ~ \left (~ 1 ~-~ {{6}\over{r}} ~+~ {{8a}\over{r^{3/2}}}
     ~-~ {{3a^2}\over{r^2}} ~ \right ) ~~
\label{eq:kappa}
\end{equation}
and is zero at the marginally stable orbit.  For a Schwarzchild black hole,
$\kappa$ reaches a maximum at $r=8$ and is approximately $71$ Hz for $M =
10~M_\odot$.  Note that for a fixed $M$, the maximum $\kappa$ for $a \sim
1$ is approximately equal to $\Omega$ at the marginally stable orbit for $a
\sim 0$.  This coincidence of time scales has led to alternative suggestions
that the 300 Hz feature in GRO~J1655-40 can be attributed to either the
Keplerian frequency at the marginally stable orbit or to the maximum
epicyclic frequency, depending upon whether one adopts $a \sim 0$ or $a
\sim 1$, respectively, for this source.

Just as the radial epicyclic frequency differs from the Keplerian orbital
frequency in General Relativity, so does the vertical epicyclic frequency
for a spinning black hole [\cite{kato90}, \cite{kato93}, \cite{perez97}].
The vertical epicyclic frequency, $\Omega_\perp$, is the frequency at which
a free particle oscillates about its original circular orbit if it is given a
vertical perturbation.  It is equal to the Keplerian orbital frequency for
both Newtonian gravity and Schwarzchild black holes.  In
general, for non-zero black hole angular momentum it is given by:
\begin{equation}
\Omega^2_\perp ~=~ \Omega^2 ~ \left (~ 1 ~-~ {{4a}\over{r^{3/2}}}
     ~+~ {{3a^2}\over{r^2}} ~ \right )
\label{eq:omegaperp}
\end{equation}
As we will discuss in \S\ref{sec:cmode}, this frequency is relevant in
determining the frequencies of global ``corrugation modes'' in accretion
discs [cf. \cite{kato90}, \cite{kato93}, Perez (1993), \cite{ipser96}].

The final gravitational frequency that we need to consider is the
Lense-Thirring precession frequency [\cite{lense}]. If an azimuthally
orbiting ring of matter is tilted out of the equatorial plane, it will
begin to precess due to frame dragging effects if the black hole has
non-zero angular momentum.  This precession frequency, $\Omega_{LT}$, is
given by the expression:
\begin{equation}
\Omega_{LT} ~=~ {{2a}\over{r^3}} ~~.
\label{eq:lense}
\end{equation}
Note that for $a\sim 1$ it is on the order of $\Omega$, for radii near the
marginally stable orbit.  Furthermore, $\Omega_{LT}$ has a strong radial
dependence (cf. \S\ref{sec:lense}).

\begin{figure} 
\vspace{-0.8 true in}
\centerline{\psfig{figure=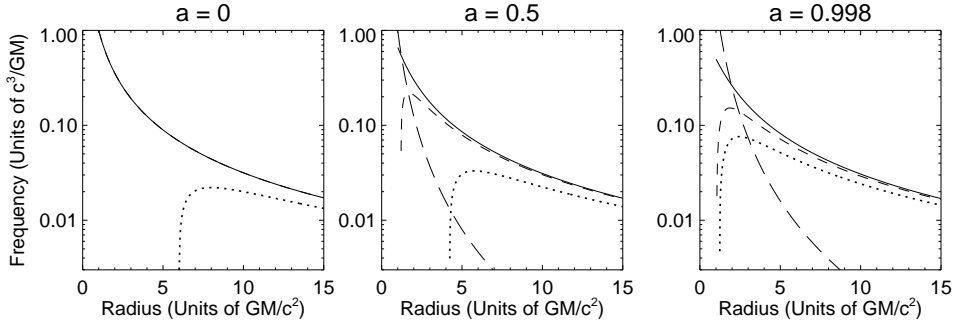,width=0.95\textwidth}}
\vspace{-0.8 true in}
  \caption{Characteristic frequencies in an accretion disc around a black
  hole, normalized to $c^3/GM ~\approx~ 3.2 \times 10^3$ Hz for $M =
  10~M_\odot$. The angular momentum parameter, $a = cJ/GM^2$, where $J$ is
  the angular momentum of the hole.  Solid line is the Keplerian frequency,
  $\Omega$, dotted line is the radial epicyclic frequency, $\kappa$, short
  dashed line is the vertical epicyclic frequency, $\Omega_\perp$, and the
  long dashed line is the Lense-Thirring precession frequency, $\Omega_{LT}$.}
  \label{fig:freqs}
\end{figure}

In Figure \ref{fig:freqs} we plot these various frequencies for three
angular momentum parameters, $a=0$, $a=0.5$, $a=0.998$.  In the inner
regions of accretion discs around black holes, these frequencies typically
range from $\sim 100-1000$ Hz, for $M=10~M_\odot$, with the highest
frequencies being achieved for $a \sim 1$.

Note that, at least for the Schwarzchild case, the Keplerian and epicyclic
frequencies can be reasonably approximated in Newtonian gravity via the
use of a ``pseudo-Newtonian'' potential.  Two such potentials that have
been used in the literature are given by:
\begin{equation}
\Phi_{NW} ~=~ - r^{-1} \left [ ~ 1 ~-~ {{3}\over{r}} ~+~ {{12}\over{r^2}} ~
     \right ] ~~,~~
\Phi_{PW} ~=~ - r^{-1} \left [ ~ 1 ~-~ {{2}\over{r}} ~ \right ]^{-1} ~~,
\label{eq:myphi}
\end{equation}
[cf. \cite{nowak91}, \cite{paz}].  Both of these potentials have $\kappa =
0$ at $r = 6$, and the $\Phi_{NW}$ potential has been used for 
Newtonian calculations of modes in relativistic discs about Schwarzchild
black holes [cf. \cite{nowak91}, \cite{nowak92}, \cite{nowak93},
\cite{nowak97}.] 

All of the above frequencies have strong radial dependences, so it is
unlikely that any one of them can lead to a narrow feature at a discrete
frequency without some mechanism that couples a range of
radii\footnote{Unless, of course, there is a ``special'' radius, such as
the marginally stable orbit radius, which picks out a discrete frequency.
However, it then may become very difficult to achieve a large luminosity
modulation.}.  Such coupling is most likely achieved via hydrodynamic
effects, i.e. pressure, viscosity, etc.  To this end, the most relevant
{\it hydrodynamical} frequencies are given by the frequency of sound (for a
given wavelength), $\omega_{cs}$, and the vertical Brunt-V\"ais\"al\"a
(i.e. buoyancy) frequency, $N_z$.  In terms of the pressure, $P$, the
density, $\rho$, and the radial and vertical perturbation wavelengths,
$\lambda_r$ and $\lambda_z$, these frequencies are given by:
\begin{equation}
\omega^2_{cs} ~=~ c_s^2 k^2 ~=~ c_s^2 ~(2 \upi)^2 ~\left ( ~ \lambda_r^{-2} ~+~
    \lambda_z^{-2} ~ \right ) ~~,
\label{eq:sound}
\end{equation}
\begin{equation}
N_z^2 ~=~ \left [ ~ \rho^{-2} {{\partial \rho}\over{\partial z}} ~-~
     (\gamma \rho P)^{-1} {{\partial P}\over{\partial z}} ~ \right ] ~
     {{\partial P}\over{\partial z}} ~~.
\label{eq:bouyancy}
\end{equation}
For a thin disc, the sound speed is of order $h \Omega$, where $h$ is the
disc half-thickness [cf. \cite{shakura73}].  For radial wavelengths of
${\cal O}(r)$, the frequency of sound waves, $\sim (h/\lambda_r) \Omega$,
is typically much less than the Keplerian frequency.  The buoyancy
frequency, as it deals with gradients on scales of ${\cal O}(h)$, can be
comparable to the vertical epicyclic frequency [cf. \cite{kato93}].
(Due to symmetry, however, the buoyancy frequency vanishes at the disc
mid-plane.)   Both of these frequencies play a part in the theories
discussed in \S\ref{sec:disko}.

\section{Theoretical Models of Disc Oscillations}\label{sec:theory}

\subsection{Historical Review}\label{sec:history}

The study of oscillations relevant to accretion discs around black holes
has a history that dates back more than 30 years.  It was quickly realized
that the $\alpha-$disc model of \cite{shakura73} shows both viscous and
thermal instabilities for high luminosities above ${\cal O}(10\% ~L_{\rm
Edd})$ [\cite{shakura76}].  These instabilities have often been invoked to
explain the intense, broad band variability seen in BHC low/hard
states\footnote{It should be noted, however, that as shown in
Fig. \ref{fig:lum} and discussed in Nowak (1995), low/hard state
luminosities are nominally below the instability limits of
\cite{shakura76}.}.

As discussed in \S\ref{sec:obshist}, until recently there has been little
observational evidence for discrete (possibly stable) periodic modes in
accretion discs about black holes.  However, as was first realized by
\cite{kato80}, not only can accretion discs support discrete oscillatory
modes, but also the effects of General Relativity modify the mode spectrum
and determine the regions in which modes can be trapped.  By considering
adiabatic perturbations of an isothermal disc, \cite{kato80} showed that the
roll-over of $\kappa$ to zero at the marginally stable orbit leads to a
cavity that effectively traps acoustic modes (i.e. $p-$modes) at the disc
inner edge (cf. \S\ref{sec:pmode}).

Later, \cite{okazaki87} considered isothermal perturbations of isothermal
discs and showed that it was also possible to trap what are essentially
internal gravity modes (i.e. $g-$modes) near the epicyclic frequency maximum
(cf. \S\ref{sec:gmode}).  Vishniac \& Diamond (1989) considered travelling
wave versions of $g-$modes that had an azimuthal dependence $\propto \exp[i
m \phi]$, with $m=1$.  These modes were invoked as a possible mechanism for
angular momentum transport in an accretion disc.  [Note, however, that the
strictly Newtonian modes considered by Vishniac \& Diamond (1989) were not
trapped modes, and furthermore they had very small vertical wavelengths,
i.e. $\lambda_z \ll h$.]

In a series of papers Nowak \& Wagoner (1991,1992,1993) adopted a
Lagrangian perturbation approach [cf. Lynden-Bell \& Ostriker (1967),
Freidman \& Schutz (1978a,b)] to the study of disc oscillations.  They
showed that the acoustic modes of \cite{kato80} and \cite{okazaki87} are
high and low frequency limits, respectively, of the same dispersion
relationship, which itself is the strong-rotation limit of the simplest
form of the ``helioseismology'' dispersion relationship
[cf. \cite{hines60}].  Hence, they gave these modes the name
``diskoseismology''.  These works used a pseudo-Newtonian potential
(eq. \ref{eq:myphi}) to mimic the effects of General Relativity; however,
they also generalized the calculations to adiabatic perturbations of
non-isothermal discs.  As discussed by Freidman \& Schutz (1978a,b), by
adopting a Lagrangian perturbation approach one can calculate a conserved
(in the absence of dissipative forces) ``canonical energy'' for the
modes\footnote{As Freidman \& Schutz (1978a,b) point out, assigning an
energy to a mode in a rotating medium is an extremely subtle issue, and
many prior works were incorrect in determining this quantity.}.  Accurately
determining the canonical energy allows one to determine the effects of
various dissipative effects, such as viscosity [\cite{nowak92}], and allows
one to consider various excitation mechanisms for the mode
[\cite{nowak93,nowak97}].  Nowak \& Wagoner (1993,1997) also
considered ways in which the modes might lead to actual luminosity
modulations (cf. \S\ref{sec:lum}).

Ipser \& Lindblom (1992) developed a scalar potential formalism for
calculating modes of rotating systems in full General
Relativity. Gradients of the scalar potential are related to the
Lagrangian perturbation vector in both the pseudo-Newtonian formalism
[Nowak \& Wagoner (1992)] and the fully General Relativistic formalism
[Perez, et al. (1997)].  \cite{perez97} used this formalism to consider
the fully relativistic version of $g-$modes.  In addition, a third class of
modes, the $c-$modes (cf. \S\ref{sec:cmode}) recently have been identified
and described with fully relativistic formalisms [\cite{kato90},
\cite{kato93}, Perez (1993), Ipser (1996).]

\subsection{`Diskoseismology'}\label{sec:disko}

The main distinguishing feature of most of the theoretical modes described
in the works cited above is that the mode frequencies predominantly depend
upon fundamental gravitational frequencies and are not strongly effected by
hydrodynamic processes if $h \ll r$.  That is, for a thin disc, changes in
density and/or pressure (which we assume to be correlated with luminosity)
may effect the physical extent of a mode or its amplitude, but they do not
greatly effect the mode's frequency.  Furthermore, the expected mode
frequencies, $\aproxgt 100$ Hz, are comparable to the gravitational
frequencies in the innermost regions of the accretion disc.

From these standpoints, the QPO features discussed in \S\ref{sec:obshist}
are {\it not} good candidates for `diskoseismic' modes.  They are of very
low frequency, and furthermore several, such as the $3-8$ Hz QPO in Nova
Muscae [Miyamoto, et al. (1994)], have been seen to vary in frequency by
fractionally large amounts.  Likewise, the low frequency ($\sim 0.1-10$ Hz)
QPO seen in both GRS~1915+105 and GRO~J1655-40 have been observed to be
very highly variable in frequency [Morgan, Remillard, \& Greiner (1997),
Remillard, et al. (1997)].

However, the $67$ Hz feature observed in GRS~1915+105 has been {\it
consistently} observed at $67\pm2$ Hz, despite the fact that this source's
luminosity has varied by factors of two or more over the epochs during
which this feature has been observed.  Although there has been only one
detection of the 300 Hz feature in GRO~J1655--40, it was consistent with
being at a steady frequency.  The 300 Hz, if attributable to a disc
time scale, indicates a phenomenon occurring very close to the disc inner
edge.  For these reasons, we identify both of these features as candidate
`diskoseismic' modes.

For pseudo-Newtonian potential calculations of diskoseismic modes, one
defines a Lagrangian perturbation vector that describes the displacement of
a fluid vector {\it relative to its unperturbed path} [cf. Freidman \&
Schutz (1978a)].  All perturbation quantities compare the {\it displaced},
perturbed fluid element to the (moving) fluid element in the unperturbed
flow.  If we take $\vec \xi$ to be the displacement vector, then the
Lagrangian variation of the velocity, $\Delta \vec v$, is given by
$\partial \vec \xi/\partial t$ [Freidman \& Schutz (1978a)].  [As in Nowak
\& Wagoner (1991), we use $\Delta$ to denote a Lagrangian perturbation.]
The Lagrangian perturbation of the density is derived from mass
conservation, and is given by $\Delta \rho = -\rho \nabla \cdot \vec
\xi$. For adiabatic oscillations, Lagrangian perturbations of the pressure
are related to Lagrangian variations of the density by $\Delta P/P = \gamma
\Delta \rho/\rho$, where $\gamma$ is the adiabatic index [cf. Freidman \&
Schutz (1978a), Nowak \& Wagoner (1991)].

Calculationally, it is somewhat easier to determine the mode structure by
defining a scalar potential [Ipser \& Lindblom (1992), Nowak \& Wagoner
(1992)], $\delta V(r,z) \equiv \delta P/\rho$, where here $\delta$ refers
to an Eulerian perturbation\footnote{An Eulerian perturbation is one that
compares the perturbed fluid to the unperturbed fluid at a {\it fixed}
coordinate [Freidman \& Schutz (1978a)].}.  The potential $\delta V$ is
seen to be the Eulerian perturbation of the enthalpy, and furthermore the
Lagrangian perturbation is related to gradients of this potential [Nowak \&
Wagoner (1992)].  If one looks for modes $\propto \exp[-i(m\phi+\sigma
t)]$, where $t$ is time and $\sigma$ is the inertial mode frequency, then
in a WKB approximation the modes satisfy a dispersion relationship of the
form
\begin{equation}
\left[\omega^2 - \Upsilon(r) \Omega^2 \right ] (\omega^2 - \kappa^2) ~=~
\omega^2 c_s^2 k_r^2 ~~,
\label{eq:disperI}
\end{equation}
where $\omega \equiv \sigma + m \Omega$ is the corotating frequency,
$\Upsilon(r)$ is a slowly varying separation function (which comes from
separating the basic fluid perturbation equation into radial and vertical
components), $\Omega$ and $\kappa$ are the Keplerian rotation frequency and
epicyclic frequency, respectively, $c_s$ is the speed of sound, and $k_r
\equiv 2\upi/\lambda_r$ is the radial wavenumber.

This dispersion relation is essentially the same as the simplest form of
the helioseismology dispersion relation, except that $\kappa$ takes the
role of the buoyancy frequency, and $\sqrt{\Upsilon(r)} \Omega$ takes the
role of the `acoustic cutoff' frequency [cf. Hines (1960)].  Nowak \&
Wagoner (1991, 1992) showed that there are two general classes of solutions
to this dispersion relation: high and low frequency.  We identify the high
frequency ($\omega^2 \aproxgt \kappa^2$) modes with acoustic $p-$modes
[cf. Kato (1980), Nowak (1991), Perez (1993)], and the low frequency
($\omega^2 \aproxlt \kappa^2$) modes with internal gravity $g-$modes
[cf. Okazaki et al. (1987), Vishniac \& Diamond (1989), Nowak \& Wagoner
(1992), Perez, et al. (1997)].  Very qualitatively, gravitational effects
set certain fundamental oscillation frequencies, which are then modified by
pressure effects.  If the pressure forces `assist' the gravitational
forces, then a high frequency is achieved.  If the pressure forces `retard'
the gravitational forces, then a low frequency is achieved. (We will return
to this qualitative notion below when we consider mode excitation,
cf. \S\ref{sec:excite}.)

The fully General Relativistic version of the diskoseismology equations
leads to a WKB dispersion relationship of a very similar form.  
As was shown in Perez, et al. (1997), for the relativistic equations a 
WKB analysis still allows approximate separation of the governing
equation into radial and vertical dependences. The radial component
of the fluid perturbations satisfies the WKB relation
\begin{equation}
\frac{d^2W}{dr^2}+\alpha^2 \left[\Psi\left(\frac{\Omega_\perp}{\omega}\right)^2
-1\right](\kappa^2-\omega^2)W=0\; ,
\label{schroed}
\end{equation}
where $\alpha(r)$ is inversely proportional to the speed of sound at the
mid-plane and $\Psi(r)$ is a slowly-varying separation function. The
eigenfunction $W(r)$ is proportional to a radial derivative of the
potential $\delta V$ (again defined as $\delta P/\rho)$ and to the radial
component of the Lagrangian fluid displacement [Perez, et al. (1997)].

From the radial WKB equation (\ref{schroed}) one can identify three classes
of modes that are trapped. The $p-$modes, defined by
$\Psi(\Omega_\perp/\omega)^2 < 1$, are trapped where $\omega^2>\kappa^2$;
$g-$modes, defined by $\Psi(\Omega_\perp/\omega)^2 > 1$, are trapped where
$\omega^2<\kappa^2$. The third class of modes (unique to the fully
relativistic treatment), the $c-$modes, are defined by
$\Psi(\Omega_\perp/\omega)^2 \cong 1$ and may in principle be trapped in
either region. [The separated equation governing the vertical component
involves a more complicated second-order linear operator which contains the
vertical buoyancy (Brunt-V\"ais\"al\"a) frequency, $N_z$.]

\begin{figure} 
\centerline{\psfig{figure=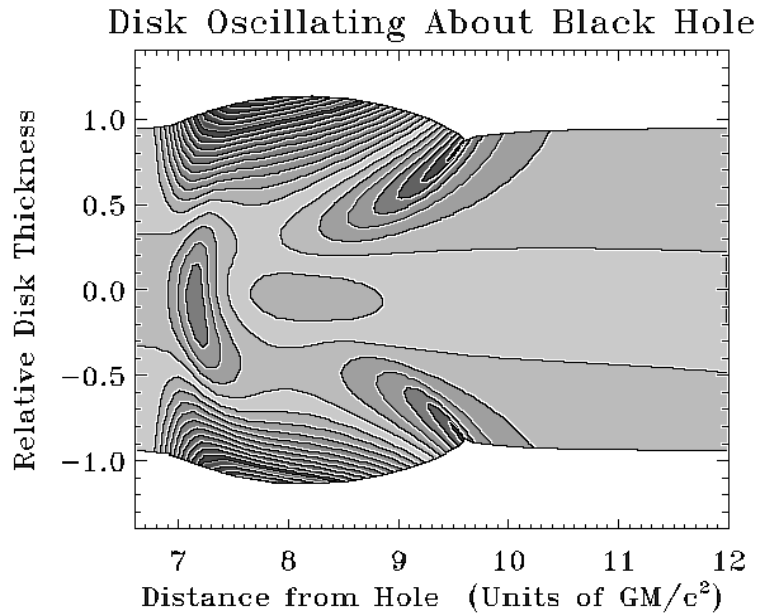,width=0.8\textwidth}}
  \caption{A cross section of a disc with an $m=0$ $g-$mode.  The disc
  thickness, as a function of radius, has been normalized to 1.  Shaded
  contours correspond to the Eulerian pressure perturbations normalized to
  the {\it local} unperturbed pressure values.  (The apparent asymmetry
  about the disc mid-plane is a numerical artifact, due to inadequate
  resolution of the contouring routine.)}
  \label{fig:gmode}
\end{figure}

The frequencies of all classes of modes are proportional to $1/M$, but
their dependences on the angular momentum of the black hole are quite
different. In principle this would allow one to measure the angular
momentum of the black hole if more than one type of mode were to be
detected in the same source. Alternatively, if one could infer the mass of
the black hole through the motion of a companion which feeds the disc, such
as has been done for GRO~J1655-40 [Orosz \& Bailyn (1997)], one could
determine the angular momentum of the black hole by a single mode
observation (given a correct identification for the class of mode
observed).  Below we briefly describe each class of mode, and we present 
representative examples in Figures \ref{fig:pmode}--\ref{fig:cmode}.

\subsubsection{P-modes}\label{sec:pmode}

\begin{figure} 
\centerline{\psfig{figure=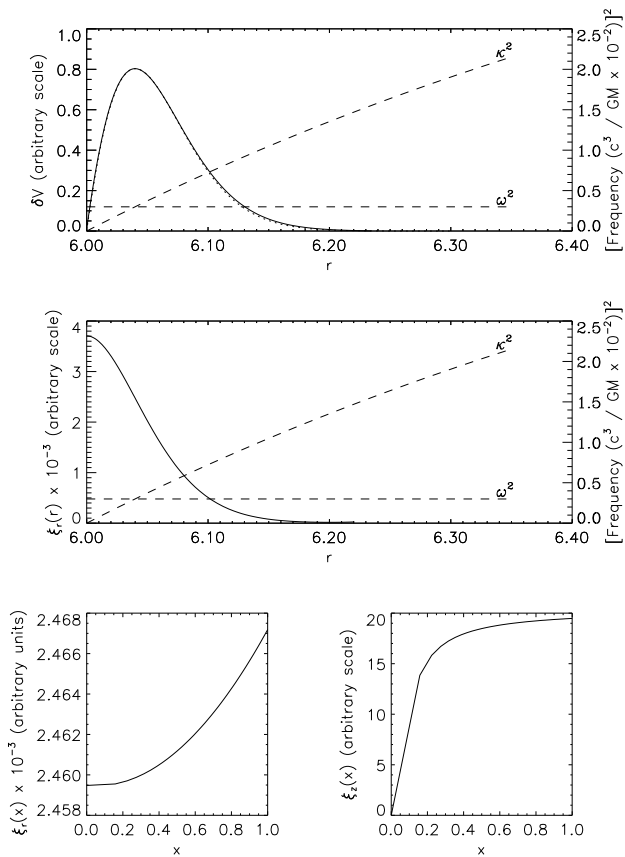,width=0.75\textwidth}}
\vskip -0.55 true in
  \caption{A representative $m=0$ $p-$mode. Here we show the Eulerian
  potential, $\delta V$, as well as the radial component of the Lagrangian
  perturbation vector as a function of radius $r$, and normalized vertical
  coordinate, $x$.  [Taken from Perez (1993).]} 
  \label{fig:pmode}
\end{figure}

As discussed by Kato (1980) and Nowak \& Wagoner (1991), $p-$modes are
trapped in regions where $\omega^2 > \kappa^2$.  For $m=0$, this leads to a
narrow trapping region ($\Delta r_{\rm mode} \ll 1$) at the disc inner edge
where $\kappa^2$ goes to zero\footnote{Trapped $p-$modes might also exist
in the large, outer region of the disc; Silbergleit 1997, Private
Communication.}.  These modes have frequencies a factor of a few less than
the radial epicyclic maximum, and tend to have $\xi_r \aproxgt \xi_z$
(i.e. radial perturbations greater than the vertical perturbations) [Nowak
\& Wagoner (1991)].  Due to their narrow confinement very near the disc
inner edge [where the `no-torque' disc boundary condition leads to very
little luminosity modulation; Shakura \& Sunyaev (1973)], we do not expect
these modes to have significant luminosity modulation.  We present an
example of a $p-$mode in Fig. \ref{fig:pmode}.

\subsubsection{G-modes}\label{sec:gmode}

Internal (gravity) modes are trapped where $\omega^2 < \kappa^2$, in the
region where $\kappa$ achieves its maximum value and where the disc is
hottest. The lowest modes can have significant vertical displacements
($\xi_z \aproxgt \xi_r$) and relatively large radial extents $\Delta r
\approx GM/c^2$. We believe that these modes will produce the greatest
luminosity modulations in the disc; thus, they may be the most observable
class of modes [Nowak \& Wagoner (1992), Perez, et al. (1997)].

To analytically approximate the eigenfunctions and eigenfrequencies of the
lowest $g-$modes, WKB solutions to the separated radial and vertical
equations of fluid perturbations can be obtained. From the symmetry of the
governing equations, it is sufficient to consider eigenfrequencies
$\sigma<0$ and axial mode integers $m \geq0$.  The resulting frequencies,
$f= -\sigma/2\pi$, of the lowest radial ($m=0$) $g-$modes are given by
\begin{eqnarray}
\label{eq:freq}
f=714\,(1-\epsilon_{nj})(M_\odot/M)\;F(a)\ \mbox{Hz}, \nonumber \\
\epsilon_{nj} \approx \left(\frac{n+\frac{1}{2}}{j+\delta}\right)\frac{h}{r}.
\end{eqnarray}
Here $F(a)$ is a known, monotonically increasing function of the black hole
angular momentum parameter $a$, with $F(a=0)=1$ and $F(a=0.998)=3.44$. The
properties of the disc enter only through the small correction term
$\epsilon_{nj}$, which involves the disc thickness $2h(r)$ and the radial
($n$) and vertical ($j$) mode numbers, with $\delta \sim 1$. Typically $h/r
\sim 0.1\,L/L_{\rm Edd}$ for a radiation-pressure dominated optically thick
disc region, where $L/L_{Edd}$ is the ratio of the luminosity to the
Eddington (limiting) luminosity\footnote{Although the frequencies of the
modes do not have a strong dependence upon $L/L_{\rm Edd}$, the modes are
no longer effectively trapped at large radii if $L/L_{\rm Edd} \aproxgt
0.3$.  For such high luminosities, the outer evanescent region for the
$g-$modes becomes narrow in radius, thereby allowing $g-$modes to
effectively couple to travelling waves in the outer regions of the
disc. The modes thus ``leak'' energy to larger radius.}. [Higher axial
$g-$modes with $m>0$ have a somewhat different dependence on $a$ than the
radial modes; Perez, et al. (1997)].

The 67 Hz feature observed in GRS~1915+105 has been associated with a
diskoseismic $g-$mode [Nowak, et al. (1997)].  If this association is
correct, then equation (\ref{eq:freq}) predicts a black hole mass of $10.6~
M_\odot$, if the hole is non-rotating, up to $36.3~ M_\odot$, if the hole is
maximally rotating. (Further aspects of this identification are discussed
in \S\ref{sec:excite}, \S\ref{sec:lum} below.)  If we identify the 300-Hz
feature observed in GRO~J1655-40 with the fundamental $g-$mode oscillating
in an accretion disc surrounding a 7.0 $M_\odot$ black hole [as
determined from spectra of the companion star; Orosz \& Bailyn (1997)],
then equation \ref{eq:freq} implies that its angular momentum is $93\%$ of
maximum.

\subsubsection{C-modes}\label{sec:cmode}

\begin{figure} 
\centerline{\hbox{{\psfig{figure=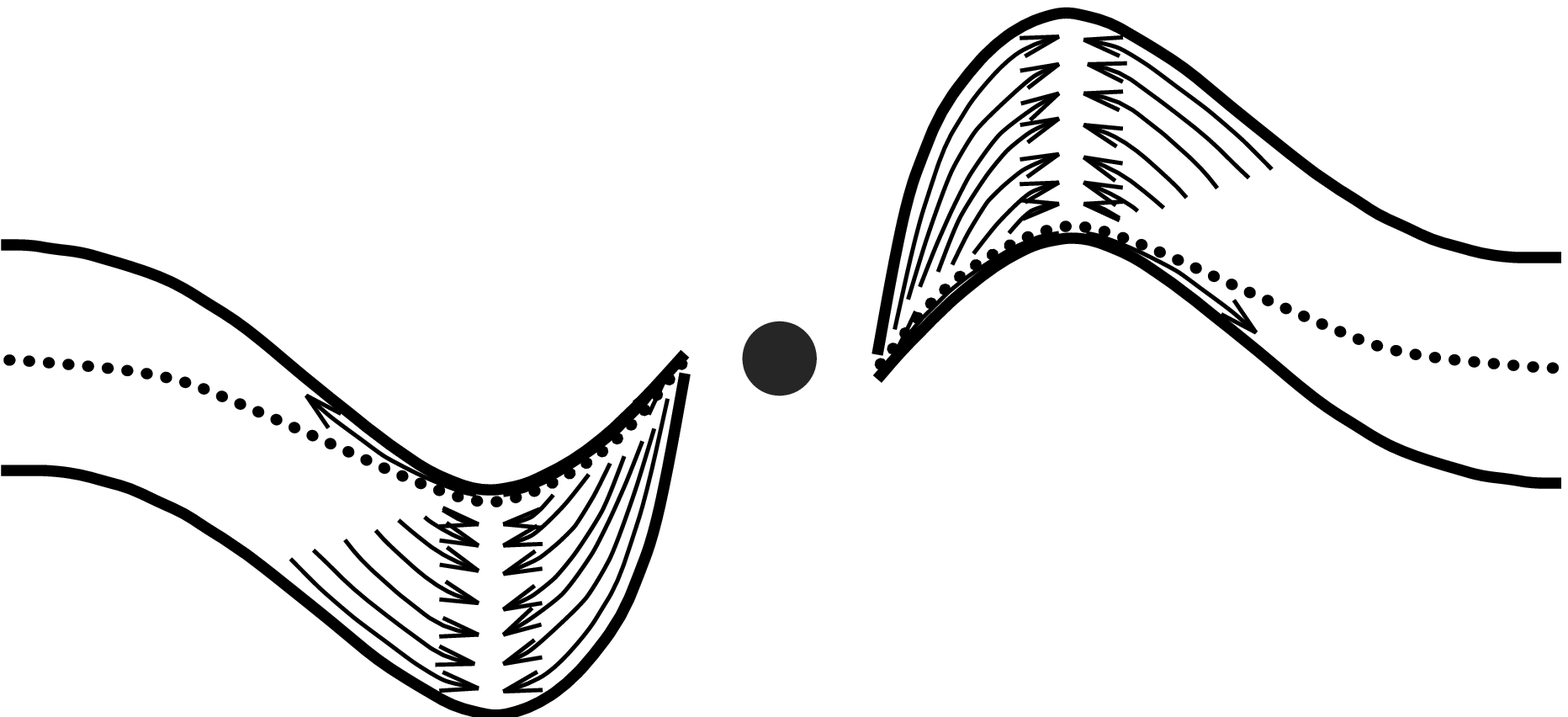,width=0.40\textwidth}}
  \hskip 0.25 true in {\psfig{figure=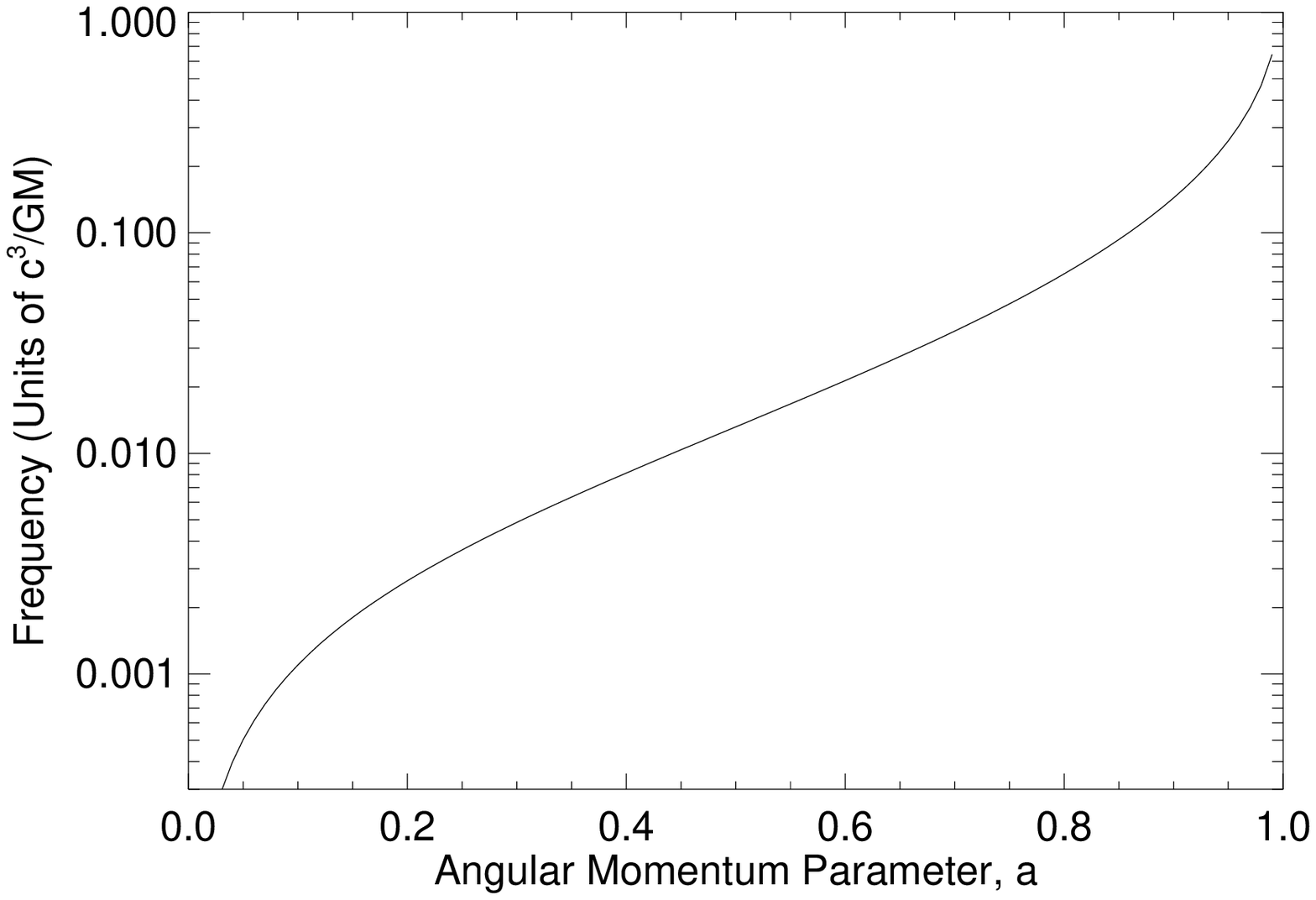,width=0.40\textwidth}}}}
  \caption{{\it Left:} Schematic drawing of a $c-$mode in an accretion disc
  (arbitrary amplitude). {\it Right:} Frequencies of the fundamental $m=1$
  $c-$mode frequency as a function of the black hole angular momentum
  parameter $a\equiv cJ/GM^2$.  Frequencies are normalized to $c^3/GM
  ~\approx~ 3.2 \times 10^3$ Hz for $M = 10~M_\odot$.}  \label{fig:cmode}
\end{figure}

To analytically obtain the approximate eigenfrequencies for the $c-$modes,
one can solve the separated WKB perturbation equations in the regime
$\Psi(\Omega_\perp/\omega)^2 \cong 1$ [Lehr, Wagoner, \& Silbergleit
(1997), Private Communication]. Here we consider $c-$modes that are
non-radial ($m \geq 1$), nearly incompressible oscillations trapped in the
very innermost region of the disc with eigenfrequencies $|\sigma| \sim
m\Omega(r_c) -\Omega_{\perp}(r_c)$, where $r_c$ is the upper radial bound
of the mode. For $m=1$ and $|a/r^\frac{1}{2}| \ll 1$ the fundamental
$c-$mode eigenfrequency is approximately the Lense-Thirring frequency
evaluated at the radius $r_c$:
\begin{equation}
|\sigma| \cong \frac{2a}{r_c^3}.
\end{equation}
Note that for the $m=1$ mode, $r_c$ is typically no greater than 10\%
larger than the marginally stable orbit radius.  The physical structure of
this mode resembles a tilted inner disc which slowly precesses about the
black-hole spin axis (cf. Fig \ref{fig:cmode}). Since the mode is nearly
incompressible, there is little temperature or pressure fluctuation in the
disc, and hence a $c-$mode may produce very little {\it intrinsic} luminosity
modulation (cf. \ref{sec:lum}). The mode may be observable, however, since
the projected area of the inner disc changes with time.  This allows the
possibility of coronal emission being Compton reflected and modulated in
the disc inner regions.  Calculations to determine the extent of the
resulting modulation are currently being undertaken [Nowak \& Reynolds
(1997), Private Communication].

Note that GRO~J1655-40 does have a high inclination, $\sim 70^\circ$, to
our line of sight [Orosz \& Bailyn (1997)] making this a potentially
promising mechanism for producing the observed 300 Hz feature.  (Also,
Compton reflection features would peak near $\sim 30$ keV, which is also
consistent with the observations.)  If we associate the 300 Hz feature with
a $c-$mode in a disc about a $7~M_\odot$ black hole, then $a \approx 0.8$
for this source.  Note that this is different from both the prediction made
from assuming a $g-$mode in this source, as well as the prediction of Cui,
et al. (1998) who associated this feature with Lense-Thirring precession
at a single radius (cf. \ref{sec:lense}).  In Fig. \ref{fig:cmode} we
show the $c-$mode frequency as a function of the angular momentum parameter
$a$.

\subsubsection{Excitation and Damping Mechanisms}\label{sec:excite}

It is possible to use a parameterized stress tensor to estimate the effects
of turbulent viscosity on the $g-$modes [Nowak \& Wagoner (1992,1993)]. The
canonical energy of a radial mode is $E_c\sim
\sigma^2\rho(\xi_z^2+\xi_r^2)~dV$, where here $dV$ is the volume
occupied by the mode. Isotropic turbulence produces a rate of change
$dE_c/dt \equiv -E_c/\tau$, with $\tau \sim |~\alpha \sigma
~[h^2/\lambda_r^2+h^2/\lambda_z^2]~|^{-1}$. Here $\lambda_r$ and $\lambda_z$,
respectively, are the radial and vertical mode wavelengths, and $\alpha$
is the standard Shakura-Sunyaev $\alpha-$parameter [Shaukura \& Sunyaev
(1973)].  The corresponding quality factor is given by
\begin{equation}
{Q_{jn}}^{-1} ~=~ (|\sigma|\tau)^{-1} ~\sim~ \left [j^2+(h/r)n^2 \right ]
     ~\alpha \; ,
\label{Q}
\end{equation}
as $\lambda_z\sim h/j$ and $\lambda_r\sim \sqrt{hr}/n$, where $j$ and
$n$ are of order of the number of vertical and radial nodes in any particular
eigenfunction. Thus, for $\alpha \ll 1$, we can have high mode $Q$
[cf. Nowak \& Wagoner (1993); Nowak, et al. (1997)].

The above estimates are for {\it isotropic} viscosity.  If the turbulence
does not efficiently couple to the vertical gradients of the modes, then
the mode $Q$ value is increased by a factor $\sim (j \lambda_r/h)^2$
[cf. Nowak \& Wagoner (1993)].  Aside from damping modes, turbulence can
also potentially excite modes.  Velocity perturbations in the disc, $\delta
\vec v$, are made up of a mode component, $\delta \vec v_M$, and a
turbulent component, $\delta \vec v_T$. Viscous damping arises from terms
of the form $\delta {v_M}_i \delta {v_T}_j$, while mode excitation arises
from terms of the form $\delta {v_T}_i \delta {v_T}_j$ [Nowak \& Wagoner
(1993)].  It is possible to make simple estimates of the magnitude of the
turbulent excitation, and balance this against the turbulent damping [Nowak
\& Wagoner (1993)].  The modes are excited to an amplitude of $|\xi^z| \sim
\alpha (h/{\lambda_r})^{3/2} ~h$, and $\sim \alpha \sqrt{{\lambda_r}/{h}}
~h$, for isotropic and anisotropic viscosity, respectively [Nowak \&
Wagoner (1993)]. If turbulence is playing the dominant role in damping and
exciting the modes, then we have the following constraints.  For isotropic
turbulence, the $Q$ value is large only for $\alpha \ll 1$; however, this
implies a correspondingly small amplitude.  For anisotropic viscosity, a
larger mode amplitude and a higher $Q$ is achieved for a given
$\alpha$. However, in either case it is likely very difficult to achieve
amplitudes as large as required to agree with the observations of
GRS~1915+105 and GRO~J1655-40 (cf. \S\ref{sec:lum}).

Another potential excitation mechanism for $g-$modes is `negative
radiation' damping [Nowak, et al. (1997)]. As a first approximation, the
$g-$modes are taken to be adiabatic.  In reality, one expects there to be
small entropy changes due to various effects, the most notable one being
radiative losses.  If one has a radiation pressure dominated atmosphere, as
is likely in high-luminosity discs, one properly should use
\begin{equation}
\Delta P ~=~ \gamma {{P}\over{\rho}} \Delta \rho ~+~ \gamma {{P}\over{s}}
     \Delta s ~~,
\label{nonadiabatic}
\end{equation}
where $s$ is the specific entropy.  The effects of 
this term for a radiation pressure dominated
atmosphere can be approximated by writing an `effective' adiabatic index,
$\gamma'$, to be substituted into the relationship between the Lagrangian
variations of the pressure and density.  Nowak, et al. (1997) showed this
term to be
\begin{equation}
\gamma' ~\approx~ \gamma \left (  1 ~+~ i {{\gamma c}\over{ 4 \omega
     \tau_{es} h}} \right ) ~\equiv~ \gamma \left ( 1~+~ i \alpha' \right )
      ~~,
\label{gammaprime}
\end{equation}
where $\tau_{es}$ is the scattering depth of the disc and $h$ is the disc
half thickness.  In the above, our ignorance of the disc's true vertical
structure is subsumed into the parameter $\alpha'$, which one expects to be
of ${\cal O}(\alpha)$ for a radiation pressure-dominated disc.

It is the imaginary component of the effective adiabatic index that leads to
radiative damping for $p-$modes but conversely leads to excitation of the
$g-$modes. One can see this by perturbing the approximate dispersion
relationship $\omega^2 \approx \kappa^2 \pm c_s^2 k_r^2$, where the $+$ is
for $p-$modes and the $-$ is for $g-$modes.  Perturbing the frequency and
the sound speed together, we have:
\begin{equation}
\delta \omega \omega ~\approx~ \pm \delta c_s c_s k_r^2 ~~,
\end{equation}
where $\delta c_s^2 \sim i \alpha' c_s^2$.  Thus, the imaginary component
of $\gamma'$ leads to damping of $p-$modes and excitation of
$g-$modes. Qualitatively, as the $p-$mode compresses gas, radiation leaks
out, thereby providing less restoring force from the compressed region, and
therefore leads to damping.  As discussed above, the $g-$mode is in some
sense opposite as the pressure works {\it against} the gravity.  Hence,
radiative leaking leads to less effective pressure restoring forces
and thereby to $g-$mode growth.

\subsubsection{Luminosity Modulation}\label{sec:lum}

For most simple $\alpha$-models, the energy generation rate per unit volume
in the accretion disc is approximated by $\alpha P(r,z) \Omega(r)$.  The
local pressure is modulated by $p-$ and $g-$modes (and to a much lesser
extent by $c-$modes); thus, there is the possibility that these modes can
lead to an observable luminosity modulation.  Furthermore, the modes not
only perturb the pressure, but they also perturb the locations of the disc
boundaries.  As discussed in Nowak, et al. (1997), the integrated variation
of the energy generation rate is therefore
\begin{eqnarray}
\delta L &\sim& 2 \pi \Bigg [
\int_{\ri+\xi_r(\ri)}^{\ro+\xi_r(\ro)} r' dr'
\int_{-z_0+\xi_z(-z_0)}^{z_0+\xi_z(z_0)} \alpha \Omega P'(r',z') dz'
\nonumber \\ &\mbox{}& - \int_{\ri}^{\ro} r dr ~ \int_{-z_0}^{z_0} \alpha
\Omega P(r,z) ~dz \Bigg ] ~ ,
\label{deltaL}
\end{eqnarray}
where $P'(r',z') \equiv P(r,z) + \Delta P(r,z)$, $r' \equiv r+
\xi_r(r)$, $z'=z+\xi_z(z)$, and $\ri$, $\ro$, $z_0$ are the disc boundaries
($\ro$ can be taken to go to $\infty$ without loss of generality).  

The $g$-modes have the most promising combination of pressure modulation
and physical extent, which makes them the most likely candidates to be
observed by direct luminosity modulation of this sort.  However, for
standard disc models, only a few percent of the total luminosity is
generated in the region of the disc where the $g-$modes can exist.  Thus,
we expect only of ${\cal O}(1\%)$ rms modulation due to the $g-$modes.
This is consistent with the observations of GRS~1915+105 and GRO~J1655-40;
however, as shown in Nowak, et al. (1997), even this modest luminosity
modulation requires large mode amplitudes.  Specifically, $\xi_z \sim h$ is
required\footnote{Note, however, that Nowak, et al. (1997) only considered
Schwarzchild black holes.  As a disc about a Kerr black hole releases a
fractionally greater amount of its energy in the $g-$mode region, one may
only require a mode amplitude a factor of several smaller.}.

Qualitatively, as the $g-$modes exist in the inner, hotter disc regions,
one expects the modulated luminosity to be harder than the average {\it
disc} luminosity.  However, both the 67 Hz and 300 Hz QPO seem to be
strongest in the {\it power law tail}, which is problematic for the
$g-$mode interpretation of these modes if one associates the power law tail
with a corona distinct from the accretion disc.

On the other hand, although one expects the $c-$modes to have relatively
little luminosity modulation due to pressure fluctuations, these modes may
be effective at modulating reflected emission.  If the power law tail is
due to a corona above a cold accretion disc, then part of this power law
will be Compton reflected [Magdiarz \& Zdziarski (1995)], and hence may be
modulated by the warping associated with the $c-$modes.  Furthermore, one
expects the reflected emission to peak in the $20-30$ keV range [Magdiarz
\& Zdziarski (1995)], which is consistent with the observations of both
GRS~1915+105 and GRO~J1655-40.

\subsubsection{Alternative Models}\label{sec:lense}

The diskoseismology $g-$mode interpretation of the 67 Hz and 300 Hz has several
points in its favor.  First, it reproduces the correct time scales.  Second,
the $g-$mode frequencies, like the observed 67 Hz frequency, are expected to
be relatively insensitive to luminosity fluctuations.  Third, mode $Q$ of
${\cal O}(20)$ is achievable with a reasonable $\alpha-$parameter. Fourth,
there is an identified mechanism (negative radiation damping) for exciting
the modes.  Fifth, the $g-$modes are capable of modulating the luminosity
of the disc, possibly as much as the observed $\sim 1\%$.  Furthermore, we
qualitatively expect the modes to be harder than the mean disc spectrum.

The are two major arguments against the $g-$mode interpretation.  First,
one requires a large mode amplitude (although this might be reduced
slightly if one considers near maximal Kerr holes).  Second, the observed
QPO features appear strongest in the power law tails, which are not
necessarily associated with a disc component.  Aside from the $c-$modes
discussed above, there are several other alternative hypotheses to the
$g-$modes.  

One possibility is fluctuations in the location of the sonic point [Honma,
Matsumoto, \& Kato (1992)], which can lie just inside of the marginally
stable orbit, where disc orbits are going from circular orbits to near
free-fall trajectories. As discussed by Honma, Matsumoto, \& Kato (1992),
the time scales associated with this `pulsational instability' of the sonic
point are fairly rapid.  Again, one needs to show that sufficient
luminosity modulation is possible with such a mechanism.

Milsom \& Taam (1997) have numerically shown that for some accretion disc
models, one finds acoustic $p-$modes that, although not trapped, peak at
the location of the epicyclic frequency maximum\footnote{Qualitatively, one
can understand this by considering the $p-$mode dispersion relationship,
which is $(\omega^2 -\kappa^2) \approx c_s^2 k^2$.  The $p-$mode group
velocity is seen to go to zero at $\omega^2 = \kappa^2$.  Thus $p-$modes
with the maximum epicyclic frequency ``stall'' at the location of the
epicyclic frequency maximum.}.  Furthermore, their frequency is that of the
epicyclic frequency maximum. Many predictions of the $p-$mode hypothesis,
such as the mass and angular momentum of the central object, are identical
to those made by the $g-$mode hypothesis.  As for the $g-$modes, it is also
difficult to achieve ${\cal O}(1\%)$ luminosity modulation, and likewise it
may also be difficult to produce features as hard as those observed in
GRS~1915+105 and GRO~J1655-40.

More recently, Cui, Zhang, \& Chen (1998) have proposed that the observed
QPO features are associated with Lense-Thirring precession.  Specifically,
these authors choose the QPO frequency to be the Lense-Thirring frequency
at the radius at which the accretion disc effective temperature is maximum,
and thus derive $a \sim 1$ for both GRS~1915+105 and GRO~J1655-40.  There
are a number of objections to this picture.  First, choosing the radius to
be that of the effective temperature maximum is somewhat arbitrary, and one
might expect that the radius at which $r^2 F_\gamma$ is maximized, where
$F_\gamma$ is the disc photon flux over the observed energy bands, is a
more natural choice.  Second, neither $F_\gamma$ nor $r^2 F_\gamma$ are
sharply peaked functions, therefore it is extremely unlikely that one can
generate a feature with $Q\sim20$, as was observed for the 67 Hz feature in
GRS~1915+105.  [From this point of view, the $c-$modes discussed in
\S\ref{sec:cmode} are the natural, `global' features to be associated with
Lense-Thirring precession.] Third, due to frame dragging effects in the
presence of viscous forces, one expects that the inner region of the disc
will be flattened and constrained to the equatorial plane on a time scale of
${\cal O}(1$s$)$ [Bardeen \& Petterson (1975)]\footnote{Cui, Zhang, \& Chen
(1998) mistakenly claimed that this time scale was very long; however, they
were quoting the time scale for the black hole's angular momentum to align
with the binary orbital angular momentum (which is of order millions of
years).  The time scale for the inner region of the disc to flatten into the
orbital plane is significantly shorter.  Note that the $c-$modes should
also be susceptible to damping by the `Bardeen-Petterson' effect; however,
as they are global modes we expect the damping time scale to be slightly
longer than that suggested by Bardeen \& Petterson (1975).}.  Fourth, Cui,
Zhang, \& Chen (1998) do not identify any luminosity modulation mechanism.
Finally, Cui, Zhang, \& Chen (1998) do not demonstrate that there is a
viable excitation mechanism for the modes [although they suggest that the
radiative warping mechanism of Pringle (1996) may be at work].  However, as
discussed above, the $c-$modes, which are qualitatively similar to the
Lense-Thirring precession suggested by Cui, Zhang, \& Chen (1998), may yet
provide a viable explanation for the observed high frequency QPO features.

\section{Summary} 

As was discussed in \S\ref{sec:history}, the study of stable oscillations in
accretion discs has a long and rich history.  However, it was not until the
advent of {\it Ginga} in the late 1980's and of {\it RXTE} in only the past
few years that this field has also become an observational one.  

Prior to the launch of {\it RXTE}, there were only a few observations of
QPO features in BHC.  Most of these features were of low frequency
$\aproxlt 10$ Hz (this was likely mainly due to instrumental limitations),
were not seen during more than one epoch, and were often broad ($Q\aproxlt
10$) and often variable in frequency.  Few theories have been put forth
that adequately describe these observations.

In the past two years since the launch of {\it RXTE}, a wealth of new
observational information has been obtained.  Again, a number of
low-frequency, often broad and highly variable features have been
observed.  Among the wealth of features seen with {\it RXTE}, two high
frequency features stand out: the 67 Hz feature in GRS~1915+105 and the 300
Hz feature in GRO~J1655-40.  The former is notable for its steady frequency,
whereas the latter is notable for its high frequency which suggests very
strongly that it comes from very close to the probable 7 $M_\odot$ black
hole in this system.

The high frequency of both features and the stability of the 67 Hz feature
suggests that these QPO might be related to stable oscillations in the
inner regions of accretion discs.  We have described a class of theories,
which we refer to as `diskoseismology', that might offer an explanation for
these observations.  The main motivations for attributing these features to 
diskoseismic modes are that these modes: 1) are related to a ``natural"
frequency in the disc (i.e. the maximum epicyclic frequency); 2) their
spectra are expected to be characteristic of the inner, hottest regions of
the disc; 3) their frequencies are relatively insensitive to changes in
luminosity; and 4) they have low rms variability.  

This latter feature, although in agreement with the observations, is the
strongest constraint.  These modes {\it cannot} be applied to systems that
show $\aproxgt 10\%$ rms variability over a wide range of energy bands.
Furthermore, it may be difficult to explain the observed spectral hardness
of the QPO.  To these ends, one needs to begin to consider more detailed
{\it dynamical} disc models that address the production of the hard
radiation.  Also, $g-$modes have been the major focus of recent study;
however, $c-$modes, which are related to Lense-Thirring precession of
warped accretion discs, may offer a better explanation of some of the data.
Finally, in all of the above work magnetic fields have been neglected.  As
shown by Balbus \& Hawley (1992), even weak magnetic fields can play an
important dynamical role.  The incorporation of magnetic fields into the
diskoseismology calculations is one of the next major steps that needs to
addressed.

Even if the features seen in GRS 1915+105 and GRO~J1655-40 do not turn out
to be diskoseismic modes, they point out two important lessons.  First,
BHC systems can produce relatively stable, high-frequency features.
Second, the {\it Rossi X-ray Timing Explorer} is capable of detecting and
characterizing these features despite their weak variability.  In a very
real sense, despite nearly twenty years of research, the study of stable
oscillations in black hole accretion systems has just begun.

\begin{acknowledgments}
The authors would like to acknowledge useful conversations with Robert
Wagoner, Mitchell Begelman, Brian Vaughan, Chris Perez, Ron Remillard, and
Ed Morgan.  M.A.N. was supported in part by an LTSA grant from NASA (NAG
5-3225). D.E.L. was supported in part by a NASA GSRP Training Grant
NGT5-50044.
\end{acknowledgments}


\begin{thebibliography}{} 

 \bibitem[Balbus \& Hawley (1992)]{bh}
	{\sc Balbus, S. A., \& Hawley, J. F.} 1992
	{A powerful local shear instability in weakly magnetized disks. I -
	Linear analysis. II - Nonlinear evolution.}
	{\it Ap.~J.} {\bf 376} 214--233.

 \bibitem[Bardeen \& Pettersen (1975)]{bard}
	{\sc Bardeen, J. M., \& Petterson, J. A.} 1975 
	{ }
	{\it Ap. J.} {\bf 195} L65.

 \bibitem[Binney \& Tremaine (1987)]{binney}
	{\sc Binney, J.,  \& Tremaine, S.} 1987
	{\it Galactic Dynamics.} Princeton Press.

 \bibitem[Chen, Swank, \& Taam (1997)]{chen97}
	{\sc Chen, X., Swank, J. H., Taam, R. E.} 1997
	{The pattern of correlated X-ray timing and spectral behavior in
	GRS~1915+105.} {\it Ap.~J.} {\bf 477}  L41--L44.

 \bibitem[Cui et al. (1997)]{cui97}
	{\sc Cui, W., Zhang, S. N., Focke, W., \& Swank, J. H.} 1997
	{Temporal properties of Cygnus X-1 during the spectral
	transitions.} {\it Ap.~J.} {\bf 484} 383--393.

 \bibitem[Cui, Zhang, \& Chen (1998)]{cui98}
	{\sc Cui, W., Zhang, S. N., \& Chen, W.} 1998
	{Evidence for frame dragging around spinning black holes in X-ray
	binaries.} {\it Ap. J.} {\bf 492} L53--L57.

 \bibitem[Davenport \& Root (1987)]{davenport}
	{\sc Davenport, W. B., Jr., \& Root, W. L.} 1987
	{\it An Introduction to the Theory of Random Signals and Noise.}
	IEEE Press.

 \bibitem[Ebisawa, Mitsuda, \& Inoue  (1989)]{ebisawa89}
	{\sc Ebisawa, K., Mitsuda, K., \& Inoue, H.} 1989 
	{Discovery of 0.08-Hz quasi-periodic oscillations from the black
	hole candidate LMC X--1.} 
	{\it P.~A.~S.~J.} {\bf	41}, 519--530. 

 \bibitem[Freidman \& Schutz (1978a)]{freidI}
	{\sc Friedman, J. L., \& Schutz, B. F.} 1978
	{Lagrangian perturbation theory of nonrelativistic fluids.}
	{\it Ap.~J.} {\bf 221} 937--957.

 \bibitem[Freidman \& Schutz (1978b)]{freidII}
	{\sc Friedman, J. L., \& Schutz, B. F.} 1978
	{Secular Instability of rotating newtonian stars.}
	{\it Ap.~J.} {\bf 222} 281--296.

 \bibitem[Grebenev, et al. (1991)]{grebenev91} 
	{\sc Grebenev, S. A., Syunyaev, R. A., Pavlinskii, M. N., \&
	Dekhanov, I. A.} 1991
	{Detection of quasiperiodic oscillations of X-rays from the
	black-Hole candidate GX 339--4.} {\it Sov. Astron. Lett.}
	{\bf 17}, 413--415.

 \bibitem[Hines (1960)]{hines60}
	{\sc Hines, C. O.} 1960
	{ } {\it Can. J. Phys.} {\bf 38} 1441.

 \bibitem[Hjellming \& Rupen (1995)]{hjellming}
	{\sc Hjellming, R. M., \& Rupen, M. P.} 1995
	{Episodic ejection of jets by the X-ray transient GRO:J1655-40.}
	{\it Nature} {\bf 375} 464. 

 \bibitem[Honma, Matsumoto, \& Kato (1992)]{honma92}
	{\sc Honma, F., Matsumoto, R., \& Kato, S.} 1992
	{Pulsational instability of relativistic accretion disks and its 
	connection to the periodic X-ray time variability of NGC 6814.}
	1992 {\it P.~A.~S.~J.} {\bf 44} 529--535.

 \bibitem[Ipser \& Lindblom (1992)]{ipser92}
	{\sc Ipser, J. R., \& Lindblom, L.} 1992
	{On the pulsations of relativistic accretion disks and rotating
	stars - the Cowling approximation.}
	{\it Ap.~J.} {\bf 389} 392--399. 

 \bibitem[Ipser (1996)]{ipser96}
	{\sc Ipser, J. R.} 1996
	{Relativistic accretion disks: low-frequency modes and frame
	dragging.} 
	{\it Ap.~J.} {\bf 458} 508--513.

 \bibitem[Iwasawa, et al. (1998)]{iwasawa98}
	{\sc Iwasawa, K., Fabian, A. C., Brandt, W. N., Kunieda, K.,
	Misaki, K., Reynolds, C. S., \& Terashima, Y.} 1998
	{Detection of an X-ray periodicity in the Seyfert galaxy
	IRAS18325-5926.} {\it M.~N.~R.~A.~S.}, in Press.

 \bibitem[Kato \& Fukue (1980)]{kato80}
	{\sc Kato, S., \& Fukue, J.} 1980 
	{Trapped radial oscillations of gaseous disks around a black hole.}
	{\it P.~A.~S.~J.} {\bf 32} 377--388.

 \bibitem[Kato (1990)]{kato90}
	{\sc Kato, S.} 1990
        {Trapped one-armed corrugation waves and QPOs.}
	{\it P.~A.~S.~J.} {\bf 42} 99--113.

 \bibitem[Kato (1993)]{kato93}
	{\sc Kato, S.} 1993
	{Amplification of one-armed corrugation waves in geometrically thin
	relativistic accretion disks.} 
	{\it  P.~A.~S.~J.} {\bf 45} 219--231.

 \bibitem[Kitamoto, et al. (1992)]{kitamoto92}
	{\sc Kitamoto, S.,  Tsunemi, H., Miyamoto, S., \& Hayashida, K.}
	1992
	{Discovery and X-ray properties of GS 1124--683 (=Nova Muscae).} 
	{\it Ap.~J.} {\bf 394}, 609--614.

 \bibitem[Kouveliotou, et al. (1992a)]{kouveliotou92a}
	{\sc Kouveliotou, C., Finger, M. H., Fishman, G. J., Meegan, C. A., 
	Wilson, R. B., Paciesas, W. S.} 1992a {\it I.~A.~U.\ Circ.} 5576.

 \bibitem[Kouveliotou, et al. (1992b)]{kouveliotou92b}
	{\sc \underbar{\ \ \ \ \ \ \ \ \ \ }.} 1992b {\it I.~A.~U.\ Circ.}
	5592.

 \bibitem[Lense \& Thirring (1918)]{lense}
	{\sc Lense, J. \& Thirring, H.} 1918
	{ } {\it Phys. Z.} {\bf 19} 156.

 \bibitem[Lynden-Bell \& Ostriker (1967)]{lbo67}
	{\sc Lynden-Bell, D. \& Ostriker, J. P.} 1967
	{On the stability of differentially rotating bodies.}
	{\it M.~N.~R.~A.~S.} {\bf 136} 293--310.

 \bibitem[Magdiarz \& Zdiarski (1995)]{mag}
	{\sc Magdziarz, P. and Zdziarski, A.} 1995
	{Angle-Dependent Compton reflection of X-rays and gamma-rays.}
	{\it M. N. R. A. S.} {\bf 273} 837--848.

 \bibitem[Milsom \& Taam (1997)]{milsom97}
	{\sc Milsom, J. A., \& Taam, R. E.} 1997
	{Two-dimensional studies of inertial-acoustic oscillations in black
	hole accretion discs.} {\it M.~N.~R.~A.~S.} {\bf 286} 358--368. 
 
 \bibitem[Mirabel \& Rodriguez (1994)]{mirabel94}
	{\sc Mirabel, \& Rodriguez, } 1994
	{A superluminal source in the galaxy.}
	{\it Nature} {\bf 371} 46.

 \bibitem[Miyamoto, et al. (1991)]{miyamoto91}
	{\sc Miyamoto, S., Kimura, K., Kitamoto, S., Dotani, T., \&
	 Ebisawa, K.} 1991 
	{X-ray variability of GX 339--4 in its very high state.}
	{\it Ap.~J.} {\bf 383} 784--807.

 \bibitem[Miyamoto, et al. (1992)]{canonicalI}
	{\sc Miyamoto, S., Kitamoto, S., Iga, S., Negoro, H., \& Terada,
	K.} 1992
	{Canonical time variations of X-rays from black hole candidates in
	the low-intensity state.} {\it Ap.~J.} {\bf 391} L21--L24.

 \bibitem[Miyamoto, et al. (1993)]{canonicalII}
	{\sc Miyamoto, S., Iga, S., Kitamoto, S., \& Kamado, Y.} 1993
	{Another canonical time variation of X-rays from black hole
	candidates in the very high flare state?} {\it Ap.~J.}
	{\bf 403} L39--L42.

 \bibitem[Miyamoto, et al. (1994)]{miyamoto94}
	{\sc Miyamoto, S., Kitamoto, S., Iga, S., \& Hayashida, K.} 1994
	{Normalized power spectral densities of two X-ray components from
         GS 1124--683.} {\it Ap.~J.}  {\bf 435} 398--406.

 \bibitem[Morgan, Remillard, \& Greiner (1997)]{morgan97}
	{\sc Morgan, E. H., Remillard, R. H., \& Greiner, J.} 1997
	{RXTE observations of QPOs in the black hole candidate
	GRS~1915+105.} {\it Ap.~J.} {\bf 482} 993--1010. 

 \bibitem[Nowak (1995)]{nowak95}
	{\sc Nowak, M. A.} 1995 {Towards a unified view of black hole high
	energy states.} {\it P.~A.~S.~P.} {\bf 718} 1207--1216.

 \bibitem[Nowak \& Wagoner (1991)]{nowak91}
	{\sc Nowak, M. A., \& Wagoner, R. V.} 1991
	{Diskoseismology: probing accretion disks I. Trapped adiabatic
	oscillations.} {\it Ap.~J.} {\bf 378} 656--664.

 \bibitem[Nowak \& Wagoner (1992)]{nowak92}
	{\sc Nowak, M. A., \& Wagoner, R. V.} 1992
	{Diskoseismology: probing accretion disks II. G-modes,
	gravitational radiation reaction, and viscosity.}
	{\it Ap.~J.} {\bf 393} 697--707.

 \bibitem[Nowak \& Wagoner (1993)]{nowak93}
	{\sc Nowak, M. A., \& Wagoner, R. V.} 1993
	{Turbulent generation of trapped oscillations in black hole
	accretion disks.} {\it Ap.~J.} {\bf 418} 187--201. 

 \bibitem[Nowak, et al. (1997)]{nowak97}
	{\sc Nowak, M. A., Wagoner, R. V., Begelman, M. C., \& Lehr, D. E.}
	1997
	{The 67 Hz feature in the black hole candidate GRS 1915+105 as a 
	possible ``diskoseismic'' mode.}
	{\it Ap.~J.} {\bf 477} L91--L94.

 \bibitem[Okazaki, et al. (1987)]{okazaki87}
	{\sc Okazaki, A., Kato, S., \& Fukue, J.} 1987
	{Global trapped oscillations of relativistic accretion disks.} 
	{\it P.~A.~S.~J.} {\bf  39} 457--473.

 \bibitem[Orosz \& Bailyn (1997)]{orosz97}
	{\sc Orosz, J. A., \& Bailyn, C. D.} 1997
	{Optical observations of GRO~J1655-40 in quiescence. I. A precise
        mass for the black hole primary.}
	{\it Ap. J.} {\bf 477} 876--896.

 \bibitem[Pazczynski \& Witta (1980)]{paz}
	{\sc Pazczynski, B. \& Witta, P.} 1980 {\it A. \& A.} {\bf 88} 23.

 \bibitem[Perez (1993)]{perez93}
	{\sc Perez, C. A.} 1993 {\it Ph.D. Thesis, Dept. of Physics,
	Stanford University.}

 \bibitem[Perez, et al. (1997)]{perez97}
	{\sc Perez, C. A., Silbergleit, A. S., Wagoner, R. V., \& Lehr,
	D. E.} 	1997
	{Relativistic diskoseismology. I. Analytical results for ``gravity
	modes''.} {\bf 476} 589--604.

 \bibitem[Press, W. H., et al. (1992)]{press92}
	{\sc Press, W. H., Teukolsky, S. A., Vetterling, W. T., \&
	Flannery, B. P.} 1992
	{\it Numerical Recipes in FORTRAN.  The Art of Scientific
	Computing, 2nd Edition.} Cambridge Press.

 \bibitem[Pringle (1996)]{pringle96}
	{\sc Pringle, J. E.} 1996 
	{\it M. N. R. A. S. } {\bf 281} 357--361.

 \bibitem[Remillard, et al. (1997)]{remillard97}
	{\sc Remillard, R. A., Morgan, E. H., McClintock, J. E., Bailyn,
	C. D., Orosz, J. A., \& Greiner, J.} 1997 
	In {\it Proceedings of the 18th Texas Symposium on Relativistic
	Astrophysics\/}
	(eds. A. Olinto, J. Frieman, \& D. Schramm) Univ. Chicago Press.

 \bibitem[Shakura \& Sunyaev (1973)]{shakura73}
	{\sc Shakura, N. I., \& Sunyaev, R. A.} 1973
	{Black holes in binary systems.  observational appearance.}
	{\it A. \& A.} {\bf 24} 337--355.

 \bibitem[Shakura \& Sunyaev (1976)]{shakura76}
	{\sc Shakura, N. I., \& Sunyaev, R. A.} 1976
	{A theory of the instability of disk accretion on to black holes
	and the variability of binary X-ray sources, galactic nuclei, and
	quasars.} {\it M.~N.~R.~A.~S.} {\bf 175} 613--632.

 \bibitem[Sunyaev, et al. (1992)]{sunyaev92}
	{\sc Sunyaev, R. A., Churazov, E., Gilfanov, M., Novikov, B., 
	Goldwurm, A., Paul, J., Mandrou, P., \& Techine, P.} 1992 
	{\it I.~A.~U.\ Circ.} 5593.

 \bibitem[Sunyaev, et al. (1993)]{sunyaev93}
	{\sc Sunyaev, R. A., et al.} 1993
	{Broad-band X-ray observations of the GRO J0422+32 X-ray nova by
	the ``Mir-Kivant'' observatory.} {\it A. \& A.} {\bf 280}
	L1--L4. 

 \bibitem[Taam, Chen, \& Swank (1997)]{taam97}
	{\sc Taam, R. E., Chen, X., Swank, J. H.} 1997
	{Rapid bursts from GRS~1915+105 with RXTE.} {\it Ap.~J.}
	{\bf 485} L83--L86.

 \bibitem[Tanaka \& Lewin (1995)]{tanaka95}	
	{\sc Tanaka, Y., and Lewin, W.~H.~G.} 1995, 
	{Black hole binaries.} 
	In {\it X-ray Binaries\/}  
	(eds. W. H. G. Lewin, J. van Paradijs, \& E. P. J. 
	van den Heuvel), Cambridge University Press.

 \bibitem[Ubertini, et al. (1994)]{ubertini94}
	{\sc Ubertini, P., Bazzano, A., Cocchi, M., La Padula, C., Polcaro,
	V.~F., Staubert, R., Kendziorra, E.} 1994
	{Hard X-ray timing observation of the Crab pulsar and Cygnus X-1.}
	{\it Ap.~J.} {\bf 421} 269--275.

 \bibitem[van der Klis (1994)]{vanderklis94} 
	{\sc van der Klis, M.} 1994
	{Similarities in neutron star and black hole accretion.} {\it
	Ap.~J.~S.} {\bf 92} 511--519.

 \bibitem[van der Klis (1995)]{vanderklis95}
	{\sc \underbar{\ \ \ \ \ \ \ \ \ \ }.} 1995
	{Rapid aperiodic variability in X-ray binaries.} 
	In {\it X-ray Binaries\/}  
	(eds. W. H. G. Lewin, J. van Paradijs, \& E. P. J. 
	van den Heuvel), Cambridge University Press.
 
 \bibitem[Vikhlinin, et al. (1994)]{vikh94}
	{\sc Vikhlinin, A., et al.} 1994
	{Discovery of a low-frequency broad quasi-periodic oscillation peak
	in the power density spectrum of Cygnus X-1 with Granat/SIGMA.} 
	{\it Ap.~J.} {\bf 424} 395--400.

 \bibitem[Vishniac \& Diamond]{vishniac89}
	{\sc Vishniac, E. T., \& Diamond, P.} 1989
	{A self-consistent model of mass and angular momentum transport in
	accretion disks.} {\it Ap.~J.} {\bf 347} 435--447. 


 

\end{thebibliography}
\end{document}